\documentclass[epj]{svjour}

\usepackage{graphicx}
\usepackage{fancyhdr}
\def\makeheadbox{{
 }}
\input{paperdef}

\def\mytitle{My title} 
\def\myauthors{My name}  
\def\mytype{My type of session}
\def\mysession{My session}

\def\mytitle{Higgs-Boson Benchmarks in Agreement with CDM, EWPO and BPO} 
\def\myauthors{S.~Heinemeyer}    
\def\mytype{Contributed Talk}    
\def\mysession{Colliders - Higgs Phenomenology}

\begin{document}

\graphicspath{{figs/}}

\title{Higgs-Boson Benchmarks in Agreement with\\ 
CDM, EWPO and BPO}
\author{S. Heinemeyer\inst{1}
\thanks{\emph{Email:} Sven.Heinemeyer@cern.ch}%
}                     
\institute{Instituto de Fisica de Cantabria (CSIC-UC), Santander, Spain
}
%
\date{}
\abstract{
We explore `benchmark planes' in the Minimal Supersymmetric Standard
Model (MSSM) that are in agreement with the measured cold dark matter (CDM) 
density, electroweak precision observables (EWPO) and $B$~physics 
observables (BPO). The \plane{\MA}{\tb}s are 
specified assuming that gaugino masses $m_{1/2}$, soft trilinear
supersymmetry-breaking parameters $A_0$ and the soft
supersymmetry-breaking contributions $m_0$ to the squark and slepton
masses are universal, but not those associated with the Higgs multiplets
(the NUHM framework).  We discuss the prospects for probing experimentally
these benchmark surfaces at the Tevatron collider, the LHC and the ILC.

\PACS{
      {14.80.Cp}{Non-standard-model Higgs bosons}   \and
      {12.60.Jv}{Supersymmetric models}
     } 
} 
\maketitle
%


\section{Introduction}

Some of the best prospects for probing the minimal supersymmetric
extension of the Standard Model (MSSM) \cite{mssm} might be
offered by searches for 
the bosons appearing in its extended Higgs sector. 
Searches at the Tevatron collider are starting to encroach significantly
on the options for heavier  MSSM Higgs bosons, particularly at large
$\tb$~\cite{CDFHiggsMSSMnew,D0HiggsMSSMnew}.  
Studies have 
shown that experiments at the LHC will be able to establish the
existence of an SM-like Higgs boson over all its possible
mass range, and also explore many options for the heavier MSSM Higgs
bosons~\cite{atlasrev,cms,CMS-TDR}. On the other hand, the LHC
might well be unable to distinguish between the lightest MSSM Higgs
boson and an SM Higgs boson of the same 
mass. The ILC would have much better chances of making such a
distinction~\cite{Snowmass05Higgs,djouadi2,deschi,ehow2,asbs2}, 
and might also be able to produce the other MSSM Higgs bosons if they
are not too heavy.
Furthermore, at least in some specific MSSM scenarios, electroweak
precision observables (EWPO) may also provide interesting
constraints~\cite{PomssmRep,ehoww} on the MSSM Higgs sector. 

In order to correlate the implications of searches at hadron colliders
and linear colliders, it is desirable to define MSSM Higgs benchmark
scenarios that are suitable for comparing and assessing the relative
scopes of different search strategies, see e.g.\
\citeres{benchmark23,cpx,sps,benchOther}.  

Since the MSSM Higgs sector is governed by the two parameters $\MA$ (or
$\MHp$) and $\tb$ at lowest order, aspects of MSSM Higgs-boson phenomenology
such as current exclusion bounds and the sensitivities of future searches
are usually displayed in
terms of these two parameters. The other MSSM parameters enter via
higher-order corrections, and are conventionally fixed according to certain
benchmark definitions~\cite{benchmark23,cpx}.
The benchmark scenarios commonly used in the literature encompass a range of
different possibilities for the amount of mixing between the scalar top
quarks, which have
significant implications for MSSM Higgs phenomenology, and also include
the possibility of radiatively-induced $\cp$ violation. 
The existing benchmark scenarios designed for the
MSSM Higgs sector are formulated entirely in terms of low-scale parameters,
i.e., they are not related to any particular SUSY-breaking scheme and
make no provision for a possible unification of the
SUSY-breaking parameters at some high mass scale, as occurs
in generic supergravity and string scenarios.
In applications of the existing benchmark
scenarios for the MSSM Higgs
sector~\cite{benchmark23,cpx}, one is normally
concerned only with the phenomenology of the Higgs sector itself.
Besides the direct searches for supersymmetric particles, other
constraints arising from EWPO, $B$-physics observables (BPO) and the
possible supersymmetric origin of the astrophysical
cold dark matter (CDM) are not usually taken into account.
This may be motivated by the fact that the additional constraints from EWPO,
BPO and CDM can depend sensitively on soft-supersymmetry breaking parameters
that otherwise have minor impacts on Higgs phenomenology. For example,
the presence of small flavor-mixing terms in the MSSM Lagrangian would
severely affect the predictions for the BPO while leaving Higgs
phenomenology essentially unchanged (see also \citere{sps} for a
discussion of this issue).

Here we review a different approach~\cite{ehhow} and adopt specific
universality assumptions about the soft
SUSY-breaking parameters, restricting our analysis of the MSSM 
to a well-motivated subspace of manageable dimensionality. 
The new \plane{\MA}{\tb}s are 
specified assuming that gaugino masses $m_{1/2}$, soft trilinear
supersymmetry-breaking parameters $A_0$ and the soft
supersymmetry-breaking contributions $m_0$ to the squark and slepton
masses are universal, but not those associated with the Higgs multiplets
(the NUHM framework) (see \citere{ehhow} for a list of
references). Within the NUHM, $\MA$ 
and $\mu$ can be treated as free parameters for any specified values of
$m_0, m_{1/2}$, $A_0$ and $\tb$, so that this scenario provides a suitable
framework for studying the phenomenology of the MSSM Higgs sector. 
Since the low-scale parameters in this scenario are derived from a small
set of input quantities in a meaningful way, it is of interest to take
into account other experimental constraints.


\section{The Benchmark planes}

The NUHM offers the attractive possibility~\cite{ehoww,ehhow} to
specify $(\MA, \tb)$ planes such that essentially the whole plane is
allowed by the constraints from WMAP and other observations~\cite{WMAP}.
This is done assuming that
$R$~parity is conserved, that the lightest supersymmetric particle (LSP)
is the lightest neutralino $\neu{1}$, and that it furnishes most of the
cold dark matter required~\cite{EHNOS}. 

The \plane{\MA}{\tb}s are defined by fixing three out of the four free
parameters, $m_{1/2}$, $m_0$, $A_0$ and $\mu$. The first two scenarios
are realized by varying $m_{1/2}$ so as to fulfill the CDM constraint.
Roughly $m_{1/2} \sim 1.2 \MA$ has to be chosen~\cite{ehhow}.
The observables that we study do not vary significantly as $m_{1/2}$ is
varied around this value. Specifically, we use the $m_{1/2}$ that gives
the value of the cold dark matter density that is closest to the central value
within the allowed range,
$0.0882 < \Omega_{\rm CDM} h^2 < 0.1204$~\cite{WMAP}.
The parameters of the first two scenarios, \Athree\ and \Afive, are
given in \refta{tab:nuhmscen}. 
The second scenario, \Afive, has been selected with a relatively small
value of $m_0$, since previous analyses of the CMSSM (where {\em all}
the scalar mass parameter are assumed to unify at the GUT scale, 
{\em not} leaving $\MA$ and $\mu$ as free parameters) indicated that values of 
$m_{1/2}$ and $m_0$ below 1~TeV are
preferred, in particular by the EWPO~\cite{ehow3,ehow4,ehoww} (see also
\citere{other}). 

\begin{table}[tbh!]
\vspace{-1em}
\renewcommand{\arraystretch}{1.2}
\BC
\begin{tabular}{|c||c|c|c|c||c|}
\cline{2-6} \multicolumn{1}{c||}{}
 & $m_{1/2}$ & $m_0$ & $A_0$ & $\mu$ & $\chi^2_{\rm min}$ \\ \hline\hline
\Athree\  & $\sim\frac{9}{8} \MA$ & 800  & 0 & 1000 & 7.1  \\ \hline
\Afive\   & $\sim 1.2 \MA$ & 300  & 0 & 800 & 3.1  \\ \hline
\Atwo\    & 500    & 1000 & 0 & 250 ... 400 & 7.4 \\ \hline
\Afour\   & 300    & 300  & 0 & 200 ... 350 & 5.6 \\ 

\hline\hline
\end{tabular}
\EC
\renewcommand{\arraystretch}{1}
\vspace{-0.5em}
\caption{
The four NUHM benchmark planes are specified by the above fixed and varying
parameters, allowing $\MA$ and $\tb$ to vary freely. All mass parameters are
in GeV. The rightmost column shows the minimum $\chi^2$ value found in
each plane at the points labeled as the best fits in the plots.
}
\label{tab:nuhmscen}
\vspace{-2em}
\end{table}

The other two \plane{\MA}{\tb}s are defined with 
fixed values of $m_{1/2}$ and $m_0$, and $\mu$ varying within a
restricted range chosen to maintain the LSP density within or below the WMAP
range. The parameters of \Atwo\ and \Afour\ are given in
\refta{tab:nuhmscen}. 
We use the $\mu$ that gives
the value of the cold dark matter density that is closest to the central value
within the allowed range, see above.

A likelihood analysis of these four NUHM benchmark surfaces, including the
EWPO $\MW$, $\sweff$, $\Ga_Z$, $(g-2)_\mu$
and $\Mh$ and the BPO 
$\br(b \to s \ga)$, $\br(B_s \to \mu^+\mu^-)$, 
$\br(B_u \to \tau \nu_\tau)$ and $\De M_{B_s}$ was performed recently
in \citere{ehoww}. The lowest $\chi^2$ value in each plane, denoted as 
$\chi^2_{\rm min}$, is shown in the rightmost column of
\refta{tab:nuhmscen}, corresponding to the points labeled as the best fits
in the plots below. We display in each of the following figures the
locations of these best-fit points by a (red) cross and 
the $\Delta \chi^2 = 2.30$ and 4.61 contours around the best-fit points 
in the \plane{\MA}{\tb}s for each
of these benchmark surfaces. These contours would correspond to the
68~\% and 95~\% C.L. contours in the \plane{\MA}{\tb}s {\it if} the
overall likelihood distribution, $\cL \propto e^{-\chi^2/2}$,
were Gaussian. This is clearly only approximately true, 
but these contours nevertheless give interesting indications
on the regions in the \plane{\MA}{\tb}s that are currently
preferred. 

We display in each plane the
region excluded (black shaded) at the 95~\% C.L.\ by the LEP Higgs
searches in the channel
$e^+e^- \to Z^* \to Z h, H$~\cite{LEPHiggsSM,LEPHiggsMSSM}.
For a SM-like Higgs boson we use a bound of $\Mh > 113 \gev$. The
difference from the nominal LEP mass limit
allows for the estimated theoretical uncertainty in
the calculation of $\Mh$ for specific values of the input MSSM
parameters~\cite{mhiggsAEC}. In the region of small $\MA$ and large
$\tb$, where the coupling of the light $\cp$-even Higgs boson to gauge
bosons is suppressed, the bound on $\Mh$ is reduced to 
$\Mh > 91 \gev$~\cite{LEPHiggsMSSM}.

The evaluation of the observables in the following section has been
performed using the code 
{\tt FeynHiggs} \cite{mhiggsAEC,feynhiggs,mhiggslong,mhcMSSMlong}. 
The new benchmark planes have been included into the code. 
This will enable the interested reader to explore the
prospects for her/his favorite experimental probe of supersymmetry in
these benchmark surfaces. More details can be found in the Appendix of
\citere{ehhow}.


\section{Phenomenological analysis}

We focus here on two of the four benchmark planes, \Athree\ and \Atwo.
The other two, \Afive\ and \Afour, respectively, show a qualitatively
similar behavior due to the same choice of fixed and varied
parameters. More information about \Afive\ and \Afour\ can be found in
\citere{ehhow}. 


\subsection{Tevatron}

\begin{figure}[htb!]
\vspace{-2em}
\begin{center}
\includegraphics[width=.4\textwidth,height=5.0cm]{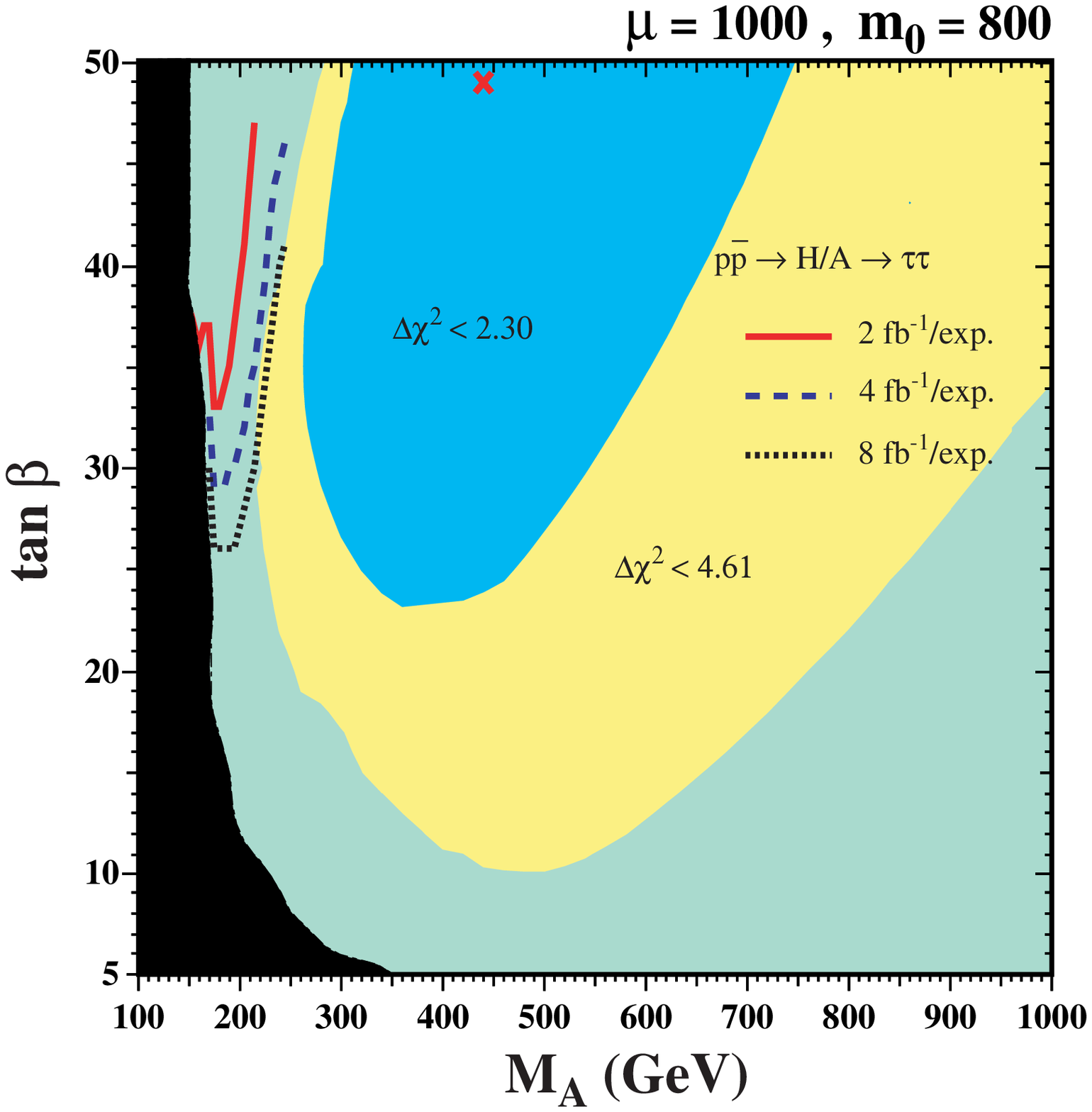}
\includegraphics[width=.4\textwidth,height=5.0cm]{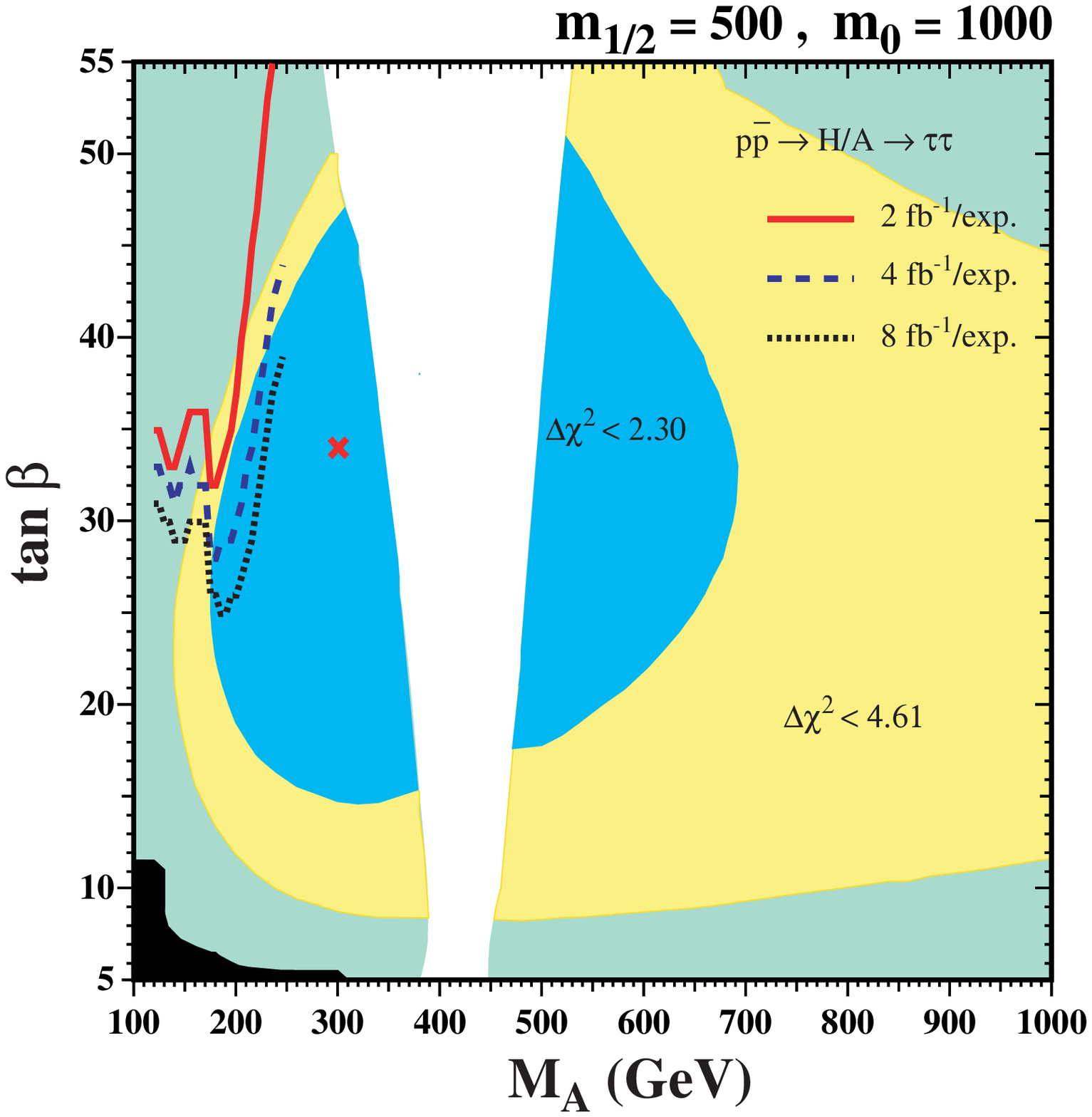}
\caption{%
The $(\MA, \tb)$ planes \Athree\ (upper) and \Atwo\ (lower plot), 
displaying the expected 95\% C.L.\ exclusion
sensitivities 
of searches for $H/A \to \tau^+ \tau^-$ at the Tevatron collider with
2, 4, 8~fb$^{-1}$ in each of the CDF and D0 experiments.
}
\label{fig:Tev02}
\end{center}
\vspace{-4em}
\end{figure}

We first consider how experiments at the Tevatron collider in the next
years could probe the
benchmark surfaces \Athree and \Atwo. We consider one
possible Tevatron signature for the MSSM Higgs sector, namely
$H/A \to \tau^+ \tau^-$, for which expectations are evaluated
using the results from~\citere{CDFprojections}. 
We see in \reffi{fig:Tev02}%
~that, at the Tevatron with 2 (4, 8)~fb$^{-1}$ of
integrated and analyzed luminosity per experiment 
the channel $H/A \to \tau^+ \tau^-$ 
would provide a 95\% C.L.\ exclusion sensitivity to $\tb \sim 35 (30, 25)$
when $\MA \sim 200 \gev$, and the sensitivity decreases slowly (rapidly)
at smaller (larger) $\MA$. In the case of the benchmark surface
\Athree, 8~fb$^{-1}$ would start accessing the region with 
$\De \chi^2 < 4.61$. 
The region $\De \chi^2 < 4.61$
could be accessed already with 2~fb$^{-1}$ in case \Atwo, and 
8~fb$^{-1}$ would give access to the region with $\De \chi^2 < 2.30$. 

\subsection{LHC}

Here we analyze the LHC reach for the heavy MSSM Higgs bosons.
In \reffi{fig:LHC02} we display in the $(\MA, \tb)$ planes the
5-$\si$ discovery contours for 
$b\bar b \to H/A \to \tau^+ \tau^-$ at the LHC, where
the $\tau$'s decay to jets and electrons or muons 
\cite{CMSPTDRjj,CMSPTDRej,CMSPTDRmj}, based on 60 or 30~\ifb\ collected
with the CMS detector.
As shown in  \citere{HNW}, the impact of the
supersymmetric parameters other than $\MA$ and $\tb$ on the discovery
contours is relatively small in this channel. The discovery contours
in the benchmark surfaces are therefore similar to each other and
to those in the ``conventional'' benchmark
scenarios~\cite{HNW}.

\begin{figure}[htb!]
\begin{center}
\includegraphics[width=.4\textwidth,height=5.0cm]{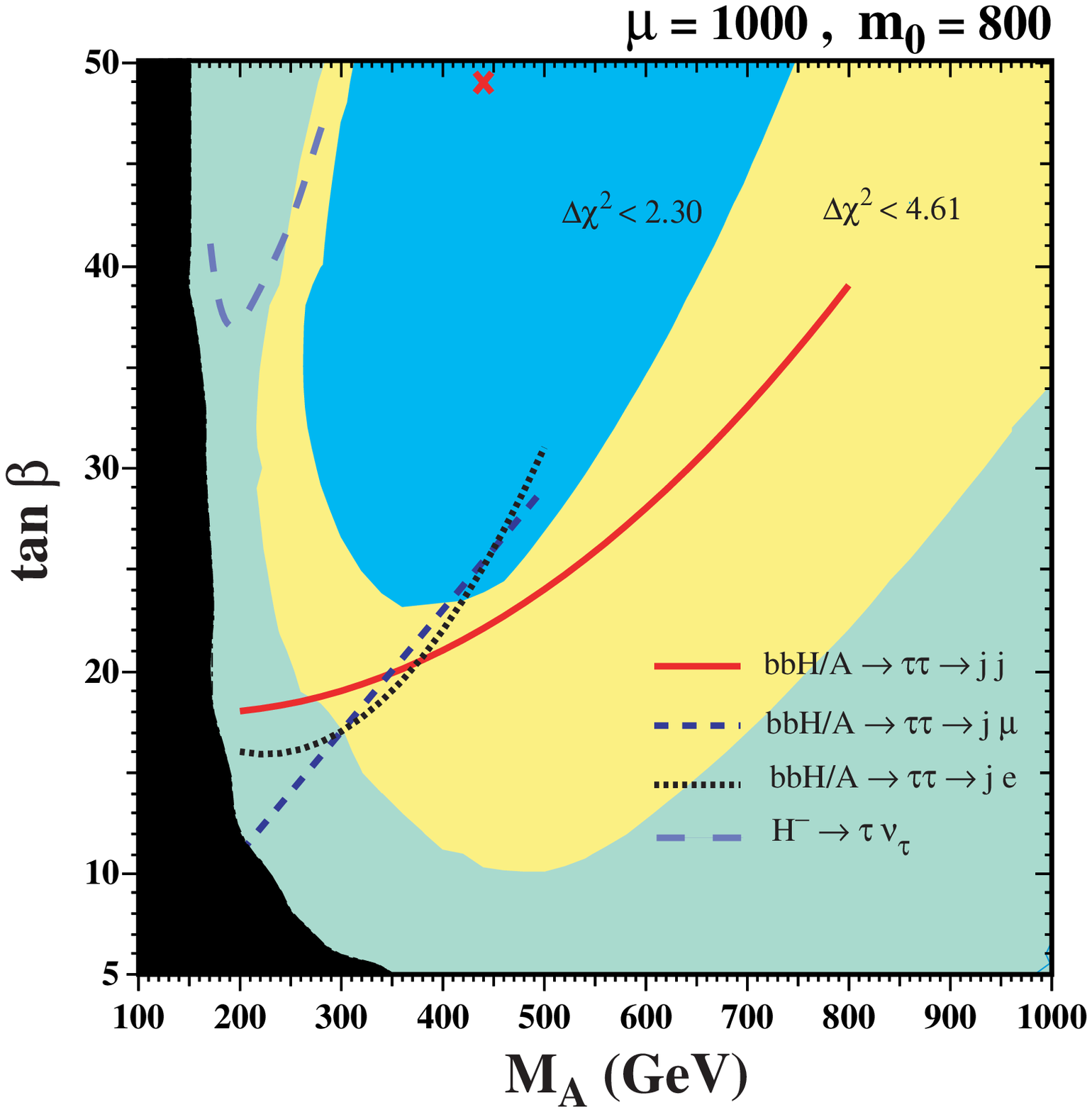}
\includegraphics[width=.4\textwidth,height=5.0cm]{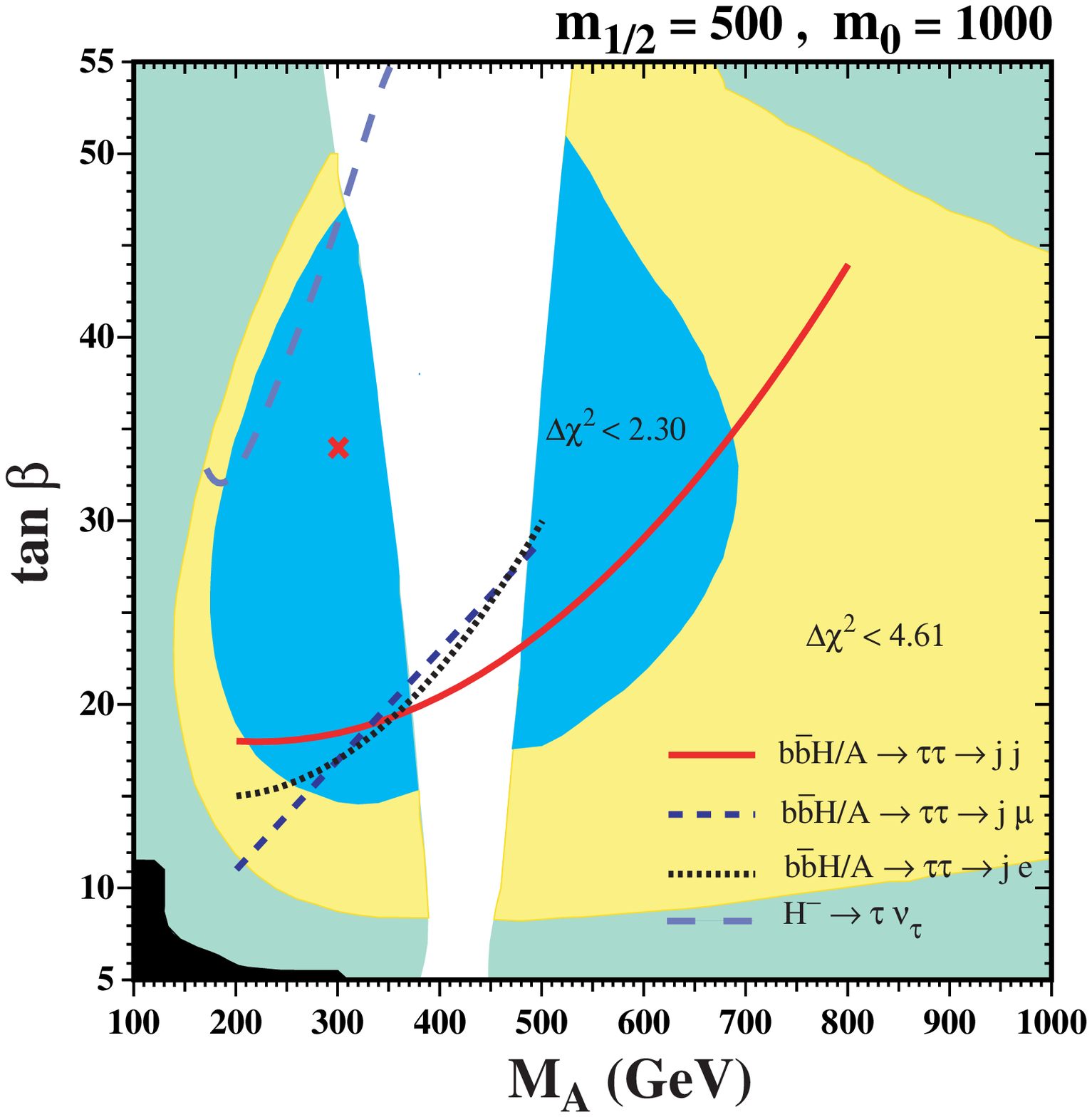}
\caption{%
The $(\MA, \tb)$ planes \Athree\ (upper) and \Atwo\ (lower plot), 
displaying the 5-$\,\si$ discovery contours 
for $H/A \to \tau^+ \tau^-$ at the LHC with 60~or 30~fb$^{-1}$
(depending on the $\tau$ decay channels) and for $H^\pm \to \tau^\pm \nu$
detection in the
CMS detector when $M_{H^\pm} > m_t$.}
\label{fig:LHC02}
\end{center}
\vspace{-1em}
\end{figure}

We also show in Fig.~\ref{fig:LHC02} the 5-$\si$ contours
for discovery of the $H^\pm$ via its $\tau^\pm \nu$ decay mode
at the LHC, in the case
$\MHp > \mt$. The coverage is
limited to $\MA < 300 \gev$ and $\tb > 30$.


\subsection{ILC}

\begin{figure}[htb!]
\begin{center}
\includegraphics[width=.4\textwidth,height=5.0cm]{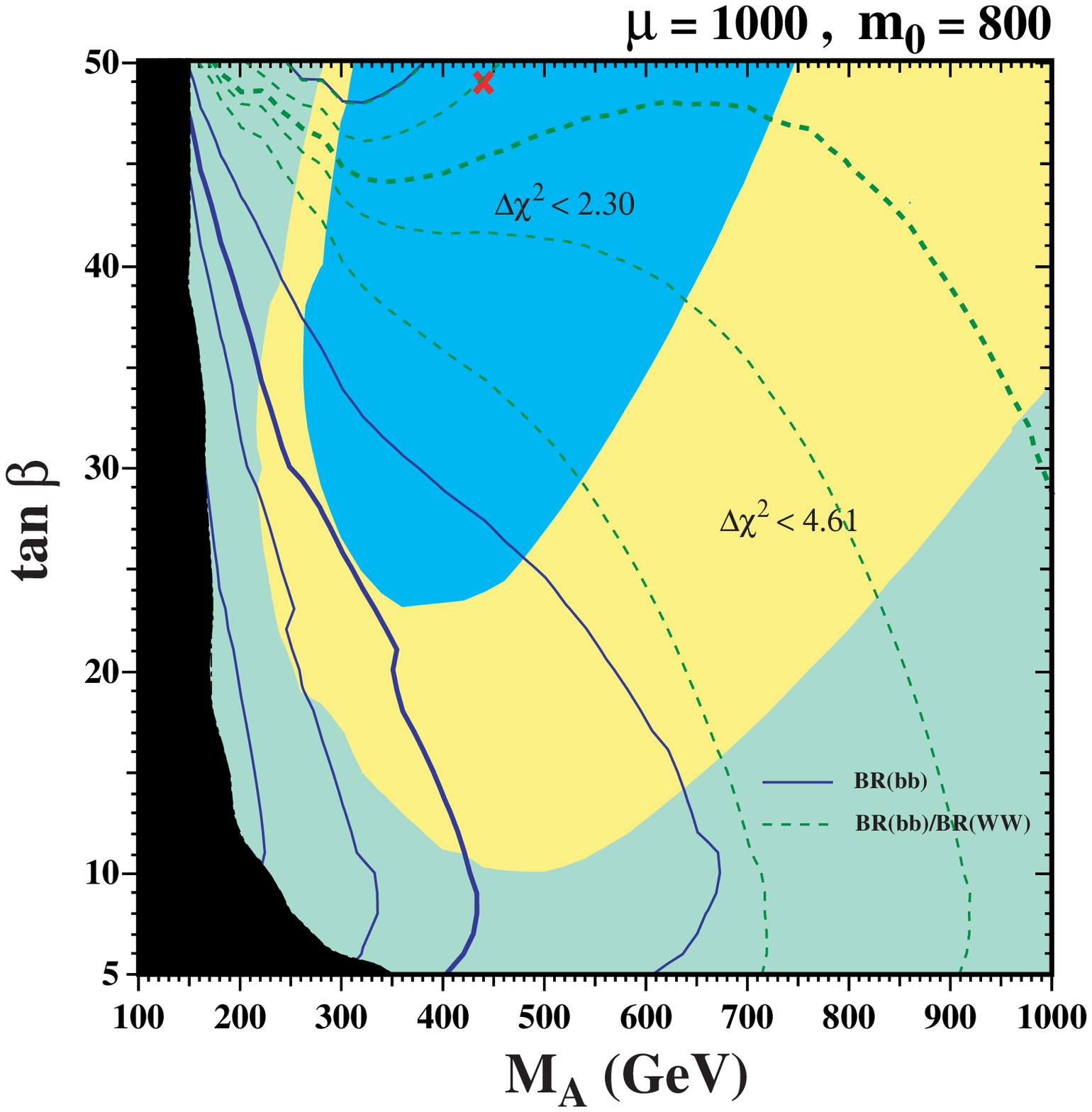}
\includegraphics[width=.4\textwidth,height=5.0cm]{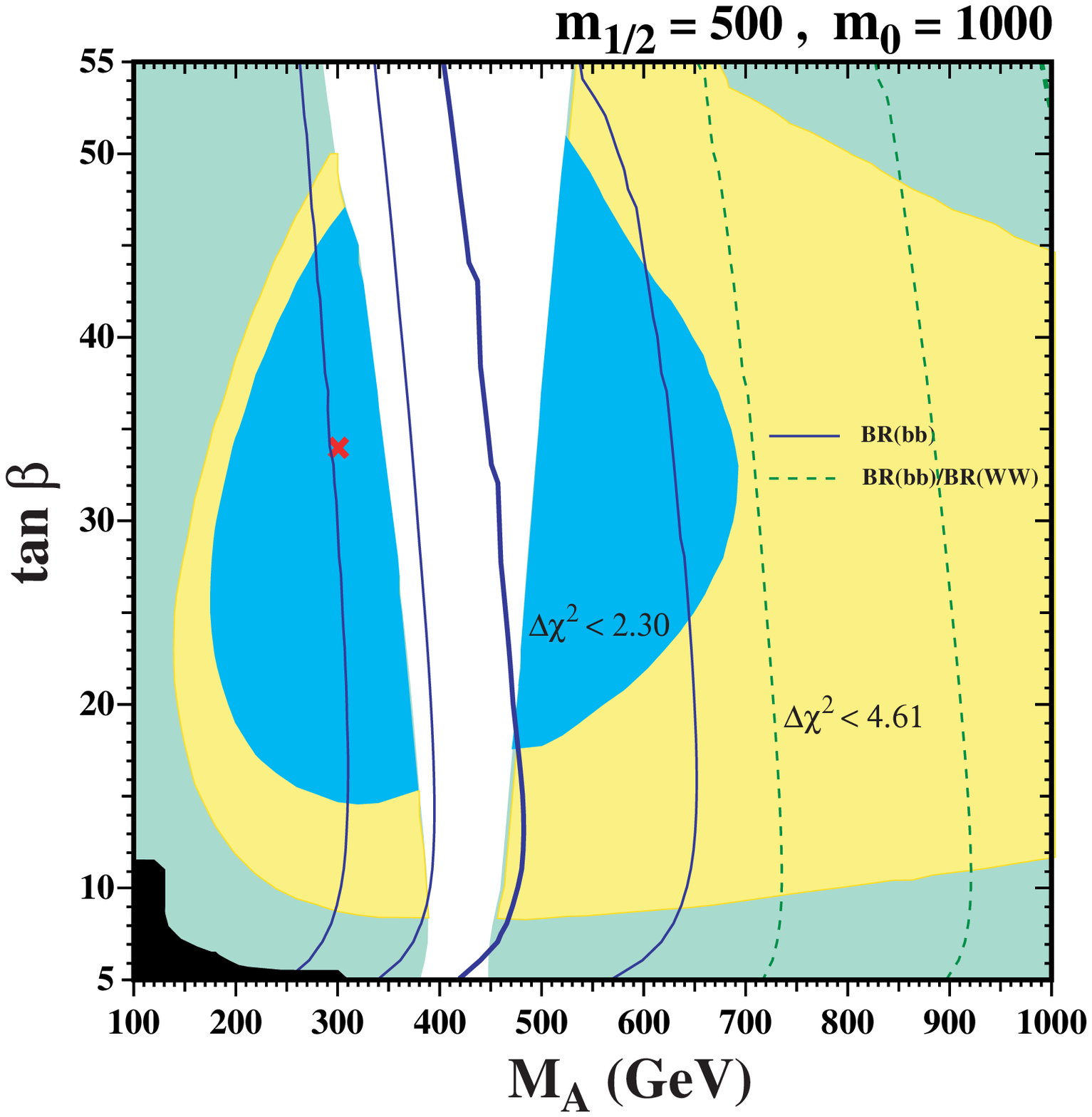}
\caption{%
The $(\MA, \tb)$ planes \Athree\ (upper) and \Atwo\ (lower plot), 
displaying $5,3,2,1,0$-$\si$ sensitivity contours 
(2-$\si$ in bold) for SUSY effects on $\br(h \to b \bar b)$ (solid blue
lines) and $\br(h \to b \bar b)/\br(h \to WW^*)$ (dashed green lines)
at the ILC. 
}
\label{fig:hbb}
\end{center}
\vspace{-2em}
\end{figure}

Finally, we show in \reffi{fig:hbb} the prospective sensitivity of an
ILC measurement of the $\br(h \to b \bar b)$ in the two
\plane{\MA}{\tb}s. The experimental precision is anticipated to be
1.5\%, see \citere{Snowmass05Higgs} and references therein.
We display as solid (blue) lines the contours of the 
$+5, +3, +2, +1, 0\,\si$ deviations (with $+2\,\si$ in bold)
of the MSSM result from the corresponding SM result.
The separations between the contours indicate how sensitively the SUSY
results depend on variations of $\MA$ and $\tb$.
Also shown in \reffi{fig:hbb} via dashed (green) lines is the
sensitivity to SUSY effects of the ILC measurement of the ratio 
of branching ratios $\br(h \to b \bar b)/\br(h \to WW^*)$, see
\citere{Snowmass05Higgs} and references therein. 
The precision measurement of the ratio $\br(h \to b \bar b)/\br(h \to WW^*)$
clearly provides a much higher sensitivity to SUSY effects than the 
measurement of $\br(h \to b \bar b)$ alone (see also \citere{deschi}). 
For nearly the full planes a $\sim\,\mbox{few}\,\si$ effect can be
established at the ILC.\\[-3em]


\subsection*{Acknowledgements}

We thank J.~Ellis, T.~Hahn, K.A.~Olive, A.M.~Weber and G.~Weiglein for 
collaboration on the work presented here.
Work supported in part by the European Community's Marie-Curie Research
Training Network under contract MRTN-CT-2006-035505
`Tools and Precision Calculations for Physics Discoveries at Colliders'.




\end{document}

\end{document}